# Estimating Bibliometric Links using Google Scholar: A Semi-Systematic Literature Mapping of Migration and Housing

Dr Boyana Buyuklieva, boyana.buyuklieva@ucl.ac.uk

Dr Juste Raimbault, j.raimbault@ucl.ac.uk

## Abstract

As the number of empirical studies increases unprecedentedly in line with expansions in higher education, theoretical developments in population studies suffer due to a discoverability crisis of related work. The systematic use of previous research is common in the medical sciences through various types of structured reviews. However, these are less common in the social sciences, despite their potential, especially in cross-disciplinary fields such as migration. We use Google Scholar to examine the niche of housing research within migration studies through a broad range of documents. The contribution of this meta-analysis is threefold. Firstly, we illustrate the association of keywords across the corpus of literature related to migration and housing and map the growth of migration literature since the 1960s. Secondly, we highlight key bridging documents using network measures. Finally, we estimate the distance in reading time between documents. Our findings suggest that the corpus of previous work at the intersection of migration and housing consists of many additive case studies, indicating a gap in integrative approaches replicable across a growing knowledge landscape.

## 1.1 Introduction

The meta-analysis in this paper provides an overview of developments in migration research with a focus on internal migration and housing. The work of Ravenstein (1889) is considered one of the seminal pieces of research and the beginning of the field of migration (Levy, Pisarevskaya and Scholten, 2020, p. 2; Pisarevskaya *et al.*, 2020, p. 457). Little over a century later, interest in the field has increased. There has been a growth in the number of journals and degrees on the subject (King, 2020, p. 4) and as well as international research networks (Levy, Pisarevskaya and Scholten, 2020, p. 20). The volume of publications and specialised outlets for these has been 'an exponential growth' since the mid-1990s (Pisarevskaya *et al.*, 2020, p. 477).

Previous bibliometric studies in migration are based almost exclusively on subscription-based academic databases such as Scopus or Web of Science. These also have different document coverage: N = 1 193 documents between the 1990s - 2018 (Piguet, Kaenzig and Guélat, 2018, p. 357), N = 2 750 documents between 1945-2015 (Nestorowicz and Anacka, 2018, p. 291), N = 21k documents between 2000-16 (Sweileh *et al.*, 2018, p. 5), N = 30k between 1959-2017 (Pisarevskaya *et al.*, 2020, p. 461) and N = 49k between 1975-2018 (Levy, Pisarevskaya and Scholten, 2020, p. 6). This paper will build on their work, adding one order of magnitude to the numeric analysis whilst focusing on a narrower search scope - the intersection of housing in migration studies in England and Wales.

## 1.2 Epistemic Communities

Epistemic communities are groups of researchers that work on a 'given subset of concepts' (Roth and Bourgine, 2005, p. 108). Where epistemic communities congregate, these form' discursive spaces' (Levy, Pisarevskaya and Scholten, 2020, p. 16) in which authors rely on similar genre knowledge to position themselves conceptually. The role and purpose of this epistemic paper trace are interesting. One perspective is that scientists want recognition for their original contributions (Merton's hypothesis after Cozzen 1981 p.3). Another is that it is not the appreciation but ascription to a broader system of academic socialising (Kaplan's hypothesis after Cozzen 1981 p.2). In this line of thought, referencing could serve as a means of persuasion such that association with another source can situate the current work in previous evidence. Referencing, therefore, is a symbolic act of embedding meaning (Cozzens, 1981, pp. 4–5). All approaches, however, converge to the result: citations create tangible links between scholarly output in any field, including migration. This tangible link can serve the purpose of quantifying the reach of a field before untangling its constituent reoccurring themes. There are several large epistemic communities within migration studies. These notably include the distinct groups that form the clusters of international and internal migration.

### 1.2.1 International vs Internal migration

Migration is a mature branch of its own within the social sciences; however, there is a 'gap' between its international and internal scope (King, 2020, p. 4). This gap has historically manifested itself as 'different literatures, concepts, methods and policy agendas' (King and Skeldon, 2010, p. 1619; Nestorowicz and Anacka, 2018, p. 284). The epistemic communities of international and internal migration scholars diverge in terms of terminologies and keyword vocabularies (Korgelli, 1994, p. 152; Nestorowicz and Anacka, 2018, p. 294). The migration dyad exists for 'good reasons' because of the issue and scope of data sources at the core of empirical studies (Nestorowicz and Anacka, 2018, p. 285).

Although international migration is what propelled migration studies, internal migration is much more common globally and, as such, 'numerically more important' (King, 2020, p. 3). Also, despite a large gap between the two approaches - both in vocabulary and inter-citation - internal and international migration are both viable strategies that can be substituted at the individual level (King, 2011, p. 137). Internal migration - as the smaller epistemic community (Nestorowicz and Anacka, 2018, p. 293) - can be informed by theoretical and empirical approaches from the former. It is, however, clear that international and internal migration are two distinct endeavours. As housing is bound to national jurisdiction, the epistemic community of greater importance is the smaller body of literature on internal migration.

## 1.3 Genre Knowledge

Different disciplines have their way of using language – this is studied in linguistics and referred to as genre knowledge, or 'intellectual scaffolds on which community-based knowledge is constructed…knowledge that professionals need in order to communicate in disciplinary communities' (Berkenkotter and Huckin, 2016, pp. 24–25). Migration exists as a topic under multiple academic disciplines, as shown in Table 1-1, adapted from comparative summaries of disciplinary approaches to migration from the edited volume by Brettell and Hollifield (2015, p. 4/24). Table 1-1 highlights key fields such as geography, economy, history and sociology to show how each area is relevant.

| Discipline | Scale of Analysis | Units of Analysis | Common Independent Variable | Common Dependent Variable | Relevance to this research |
|---|---|---|---|---|---|
| Geography | Macro, Meso | Distance, People counts | Spatial, environmental, or socio-economic geographical delinations | Migration decisions: resulting flows | Quantifying migration and delinating areas spatially |
| Demography | Micro, Meso, Macro | People counts often by groups (often ethno-racial or foreign populations) | Ethno-racial diversity, characteristics of migrant or context of reception (often in biological terms: age-structure, mortality, fertility ect.) | Size of migration flows, degree of similarity between migrants and locals | Quantifying life-course stages |
| Economics | Micro | Individual's choices (often using decision theory) | Income differentials, Human capita, demand/ supply, push/pull factors | Migration decisions: economic impact of migration flows | Framing individual choices algorithmically (e.g need thresholds and migration decisions) |
| Sociology | Macro, Meso | Groups (often by social class) | Social capital, networks or enclaves | Migrant behaviour:, including experiences societal cohesion | Framing neighbourhood characteristics (e.g community definitions) |
| History | Temporal | Timespans | Social or historical context | Migrant experience: periodisation | Setting initial conditions for housing context |
| Political Science | Macro | Population groups | Institutions, rights and responsibilities, interest groups | Policy output or outcome Political incorporation and civil engagement of migrants | Political theory, specifically definitions and evaluation of exogenous factors (e.g Land and human capital, equality and equity ) |

*Table 1-1.     Migration Across Disciplines*

Therefore, the language used by migration researchers to communicate across disciplinary communities is a point worth raising. This is because genre knowledge is also 'an individual's repertoire of situationally appropriate responses to recurrent situations' (Berkenkotter and Huckin, 2016, p. ix). Professional knowledge in this way is both an enabling factor for building on previous work and connecting with colleagues. However, it is also a limiting factor as any 'appropriate response' in one case study – bound in time and space by its empirical context - might not generalise to another.

### 1.3.1  Geography in Migration

Another important epistemic community within migration are geographers. The role of geography in the field is arguably 'under-valued' because research efforts have not been consolidated but rather dispersed across many subfields (King, 2020, pp. 1–7). Therefore, geography within migration studies has played an essential role in binding different epistemic communities: 'geographers have been perceived by our [migration] experts as the lynchpin for interdisciplinary developments' (Levy, Pisarevskaya and Scholten, 2020, p. 19).

However, as a lynchpin, the developments of geography within migration have been hallmarked as 'a descriptive discipline of facts, figures and regional monographs' - lacking 'a critical edge' (King, 2020, p. 6). This is expected when research is empirically data-constrained: the scholarship on internal migration has a geographical bias towards the global north with a dearth of contributions from Africa or Latin America (Carling, 2015). More specifically, migration scholarship is dominated by case studies of the US, British and Chinese contexts that frame the process in macro-economic terms related to urban growth, (un)employment and earnings (Nestorowicz and Anacka, 2018, p. 293). Economic geography - itself a niche within the field of economics - is also marked by a 'just do it ethos' with 'theoretical, methodological and even substantive 'turns' are announced with what seems to be increasing frequency' (Foster *et al.*, 2007, p. 1). This fickleness of the field suggests an uncoordinated and splintered approach to the spatial study of migration.

Examining publication data in the 'Geography and Environmental Studies' from the Research Assessment Exercise (RAE) of 2008 (N= 4 590 written outputs), the authors reach the same conclusion. They note that 'both physical and human geographers are promiscuous in their choice of journals' (Richards *et al.*, 2009, p. 234). The epistemic community of geographers (with the genre knowledge and skills to engage with the problem of the built environment) have consistently focused their efforts on bridging larger disciplines, although the role of housing – necessarily a spatial concern - might have fallen between the cracks. Modern migration studies have grown out of distinct fields into an interdisciplinary endeavour (Levy et al., 2020, p. 23). Thus, it is helpful to understand these developments to understand migration concerning the built environment.

## 1.4 Growth of Migration Literature

Levy et al. (2020, p. 20) note that in the period between 1975-2018, there has been a 14-fold increase in citation sources that appear only *once* in their corpus of migration literature (N = 49k). This raises concerns about the epistemic communities the average migration research might contribute to and engage with. Empirical studies are limited to specific contexts and timeframes, the nature of which can make findings difficult to generalise and thereby relate to. It is perhaps for this reason that, increasingly many outputs on migration have lower citation scores compared to works from other disciplines.

On average, one academic paper in the social sciences, more specifically migration, would cite up to 30-40 sources (Levy, Pisarevskaya and Scholten, 2020, p. 20). As more publications on migration are produced, one would also expect that studies being cited together should decrease due to the limited length of publication outlets and vast volumes of potentially relevant previous work. Consider, for example, the number of migration articles published *within demography journals alone*. These are estimated to be above 330 per year as of 2017, up from between 0-9 articles per year just over half a century ago (van Dalen, 2018)!

Most migration research is published as papers or monographs (Pisarevskaya et al., 2020, p. 457) and across many interdisciplinary journals. These types of academic output can vary in length by source. However, the average paper manuscript can be assumed to be between 25 – 40 pages of double-spaced text, including necessary data depictions (Borja, 2015). In a conservative estimate, one could consider the average research document as a paper with twenty pages of text, allowing between five and twenty for figures and references. A single page of double-spaced text is approximately 250 words (HCC Library, 2020); therefore, a conservative document length would be about 5 000 words. Assuming a normal reading speed (estimated at 225 words/minute by Duggan and Payne ( 2009, p. 15)) and no fatigue, the average research output of 5 000 words would be at least 22 minutes of reading per document. It would take at least 90 years to manually cover the corpus of papers analysed later in this paper – a long lifetime of dedicated reading with no holidays!

### 1.4.1 Literature Mapping

The science of reporting on literature – or using literature as data - has grown in the past half-century in tandem with the rise of computing and the digitalisation of research outputs (Fortunato *et al.*, 2018, p. 359).

Some milestones include the creation of the Science Citation Index (CSI) in 1961 (Cozzens, 1981, p. 16) has led to the growth of bibliometrics – 'the quantitative study' of written communication – and scientometrics – 'studies of scientific activities' such as the sociology of science through citations (Hood and Wilson, 2001, pp. 298–299). Different approaches to bibliometric analysis can be expressed as a function of rigour (reproducibility) over effort (manual or programmatic). Figure 1.1 conceptualises how there are many ways of reviewing academic literature, with a steep curve when reproducibility is taken into account. Although migration studies are a natural candidate for bibliometric analysis due to their interdisciplinary nature, there have only recently been a few precedents that have incorporated this approach (Nestorowicz and Anacka, 2018; Piguet, Kaenzig and Guélat, 2018; Sweileh *et al.*, 2018; van Dalen, 2018; Levy, Pisarevskaya and Scholten, 2020; Pisarevskaya *et al.*, 2020). This is likely because the social sciences are not as clearly indexed as the natural sciences, and migration is a relatively young field.

Reporting on literature for the critical assessment of relevant scholarship is most developed in the medical sciences. This is because controlled experiments across different groups and detailed article indexing yield themselves well for following pre-defined review protocols. Beyond the natural sciences, 'true' systematic reviews adherent to protocols are not always feasible due to the lack of two resources. Firstly, these are done in a team of at least three to include a tiebreaker when deciding on relevant articles, which is not always possible with solo research. Secondly, protocols are impossible to prescribe where there is a lack of consolidated meta-information across different databases that would allow for exhaustive and repetitive searching (Roth, 2020). In such cases, alternative approaches that involve more judgements can be employed. These are shown to the left of Figure 1.1. At the left extreme are rapid or state-of-the-art reviews that are usually done in under five weeks and serve to gather evidence on an issue (Roth, 2021b). Further to the right from these are the traditional (narrative) reviews that are characterised by unique but non-replicable descriptions and appraisals of previous work to contextualise new research (Roth, 2020). Beyond these, one begins to enter the realm of data-driven, computational approaches to assessing previous literature.

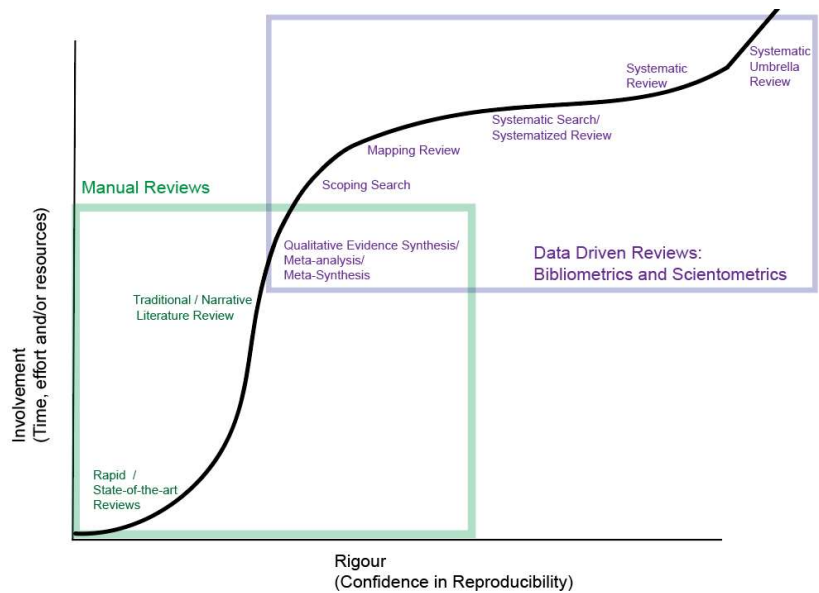

*Figure 1.1.*    *Methodological Approaches of Reporting on Literature*

At the far right of Figure 1.1 are formal 'systematic reviews', which use "accountable rigorous research methods" (Gough, Oliver and Thomas, 2017, p. 2) and involve a "minimum of two reviewers with a third to serve as a tiebreaker" (Roth, 2020). Examples of reporting on academic literature on this side of the spectrum are the family of protocols that make up the Cochrane Reviews (Higgins *et al.*, 2019) or the guidelines provided by the Preferred Reporting Items for Systematic Reviews and Meta-Analyses – PRISMA (Moher *et al.*, 2009). One level above these is 'umbrella review', or the synthesis of 'multiple reviews into one accessible and usable document' (Grant and Booth, 2009, p. 95). In a data-driven model, these would be created via an automated pipeline that stores papers matching a certain query in a database of relevant literature. One semi-automated example is the project to consolidate climate change and migration scholarship in the CLIMIG database (Piguet, Kaenzig and Guélat, 2018, p. 375; CLIMIG, 2020). At the steepest point of rigour is meta-synthesis, which can be quantitative or qualitative. What all these have in common is that they are broadly based on the researcher's judgement for numeric consistency and acceptable sensitivity of statistical results for 'meta-analyses' (Deeks, Higgins and Altman, 2019, p. 241); alternatively, judgement and interpretation of thematic compatibility and conceptual consistency for 'qualitative evidence syntheses' (Grant and Booth, 2009, p. 94; Roth, 2021a). The multitudes of methodological approaches and scope of data coverage used to understand migration (DeWind, 2020, p. 5) can make meta-synthesis on its own difficult and potentially misleading.

In such cases, 'literature mappings' can provide an alternative way for reporting on literature after it has been scoped through a narrative review. This approach aims to 'map out and categorise existing literature' (Grant and Booth, 2009, p. 94) where the field does not adhere to strict publication keyword classifications or has an inherent diversity of research outputs and, therefore, indexing. Mappings are useful to trace developments in a lineage of research and form the basis of 'systematised reviews', which provide a 'best evidence synthesis' on a thematic where systematic reviews are not possible (Grant and Booth, 2009, p. 94).

## 1.5 Methodology

This section will provide the specifications used to gather papers at the intersection of migration and housing. The corpus of bibliographic data used for this analysis consists of N = 445k documents from the openly available Google Scholar database with a temporal coverage between 1885- 2020. The choice of range begins with Ravenstein's first paper to reflect the overview of the developments within migration studies. Figure 1.2 provides a summary flow diagram of how the corpus of papers for this analysis was created. The rest of this section will explain the individual steps of the chart.

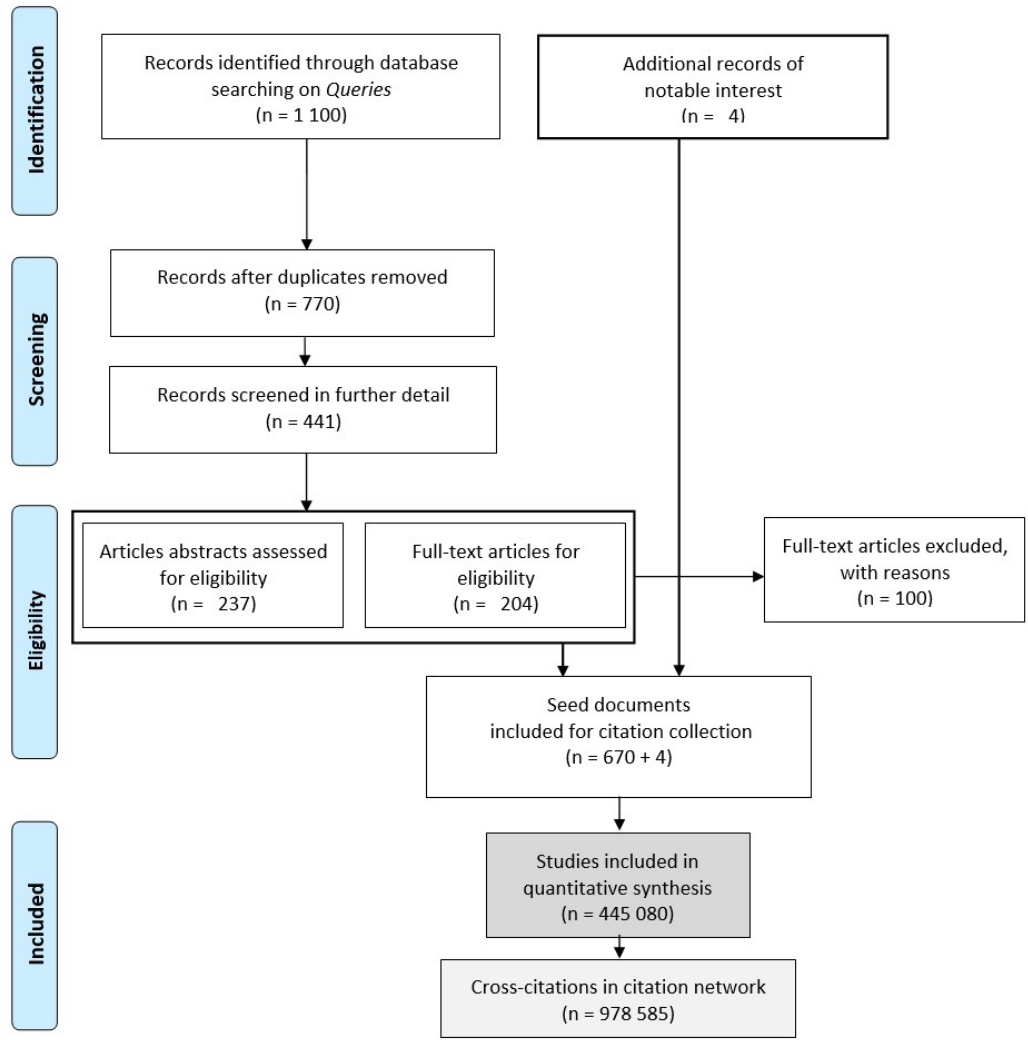

*Figure 1.2.        Flow Diagram of Papers Included following the PRISMA guide*

### 1.5.1 Google Scholar Database

The major providers of bibliographic databases are Scopus (Elsevier), Web of Science (WoS) and the freely available Google Scholar (GS) (Dhammi and Haq, 2016, p. 115). GS has broader coverage, especially in the Humanities and Social Sciences, than the other two major citation databases (Martín-Martín *et al.*, 2018, p. 20). An advantage of using the Scholar database is that it has good coverage of various disciplines and monographs, not only journal articles. The latter is an essential point because books are difficult to capture with more structured bibliometrics whilst also popular output formats for research in geography and the built environment (Richards *et al.*, 2009, p. 12). However, citation data is often bound to licensing and is collected by proprietary algorithms, which means that this is either not accessible for free as in Scopus and WoS; or open with meta-data access restrictions (Martín-Martín, Thelwall and López-Cózar, 2020, p. 6).

### 1.5.1.1 Scraping Google Scholar

Bibliographic data was co-created with Dr Juste Raimbualt. We used a scraper to emulate a manual search on GS given an anonymous request – i.e. a search was done without a user login and website cookies, as shown in Raimbault (2017, p. 5). For each query (Figure 1.3-1), the first data collection step consisted of gathering the top 100 paper results per query. Figure 1.3-2 illustrates how the top results are ordered using the GS ranking algorithm. This first step creates a scoping corpus, which underwent manual inspection to create a seed corpus of papers. After the manual sanity-check of results, the subsequent data collection steps involved gathering the top 100 documents citing each of the documents in the seed corpus (Figure 1.3-3). A complete description of the tool used for data collection collaboration is provided by Raimbault (2017) under' Database Construction'.

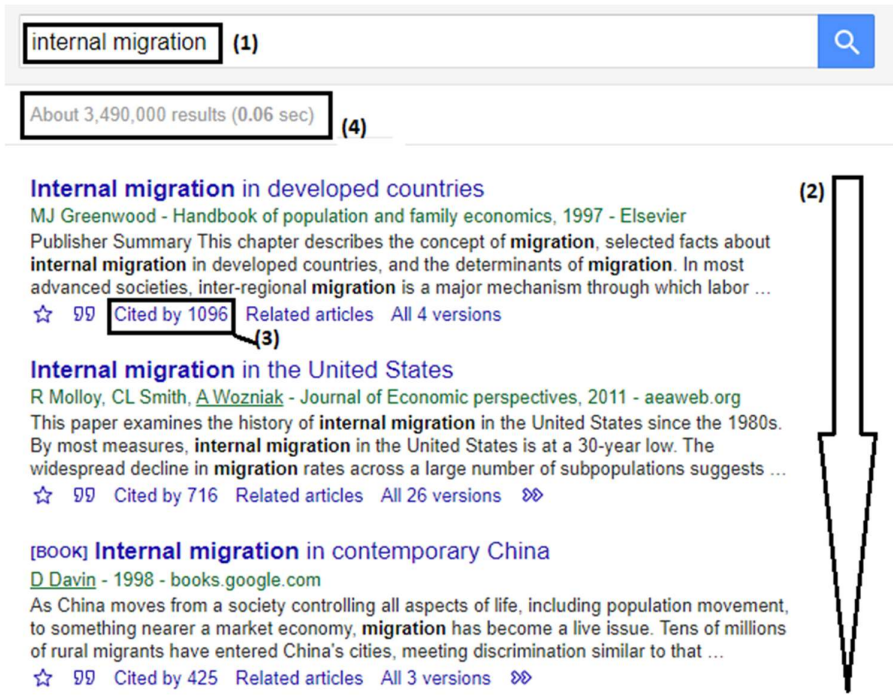

*Figure 1.3.        How Queries Relate to Papers*

Eleven queries were used (as shown in Figure 1.6) to exhaustively gather papers relevant to migration and housing and create a scoping corpus of papers for manual inspection. The large number of queries combined with the limited number of requests permitted to GS meant that collecting papers necessitated several weeks.

The most time-consuming stage was the second step of data collection – the gathering of citing papers – because of the exponential nature of links between papers. Citation collection was done in two layers (depth = 2). This means that the papers collected were those that cite the seed corpus up to two references way (see Figure 1.3-3).

The first layer was produced by including the top 100 articles that cited each of the papers within the seed corpus (shown in Figure 1.3-3). The second citations layer was created analogously by collecting the top 100 papers that cited documents in the first level. The relationship between the two layers is shown with an example in Figure 1.4. Together all documents gathered at this stage completed the citation corpus of documents for this analysis.

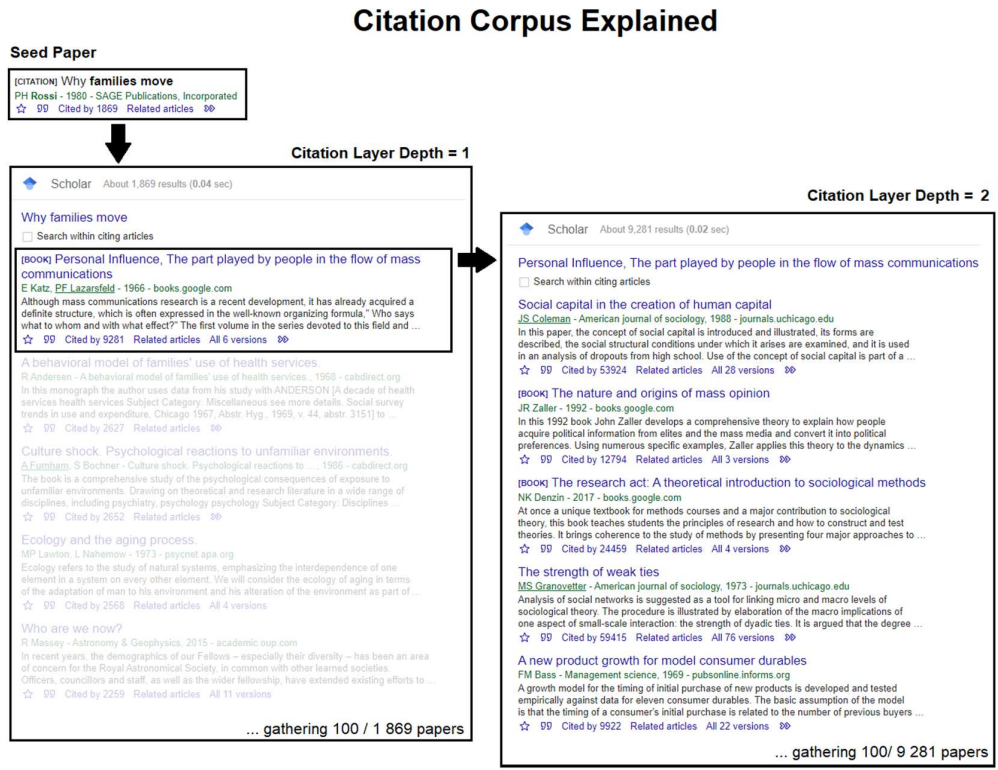

Figure 1.4        Layers of the Citation Corpus

### 1.5.2  Queries

The most important part of the methodology is the choice of queries and consistent keywords, as these ultimately define the scope of papers included in the final corpus. This next section will present the logic for choosing keywords, queries and four references of notable interest.

#### *1.5.2.1  Choice of Key Words for Queries*

It is important to note that the aim of the first stage is to be exhaustive in harvesting papers within the scope of interest. For this reason, although the GS does not have a thesaurus classification (Gibson, 2020), it was useful to consult other databases' classification of keywords. The wording of queries entries (as shown in Figure 1.3-1) was guided by a manual literature exploration and examination of various library database thesauri, notably the Humanities and Social Science Electronic Thesaurus (HASSET) compiled at the UK Data Archive (UK Data Archive, 2020).

The keywords of interest within the HASSET hierarchy of synonyms are mapped out in Figure 1.5. These show that both 'internal migration' (IM) and 'residential mobility' (RM) are narrow terms under' population migration'. Interestingly, the RM as a keyword combination has a niche belonging as a sub-term within the scholarship on internal migration. However, it also exists as a prominent theme directly under the broader scholarship on 'life histories', suggesting that RM congregates epistemic communities that add a longitudinal nuance to migration studies. Another point to note in Figure 1.5 is that RM is the gatekeeper node to terms relating to housing. It is related to 'housing needs', 'relocation expenses', and used for 'moving house' and 'housing history'. RM also carries a demography tint – due to the fact that people age over time -  with the narrow niche concerned with youth mobility and 'leaving home'.

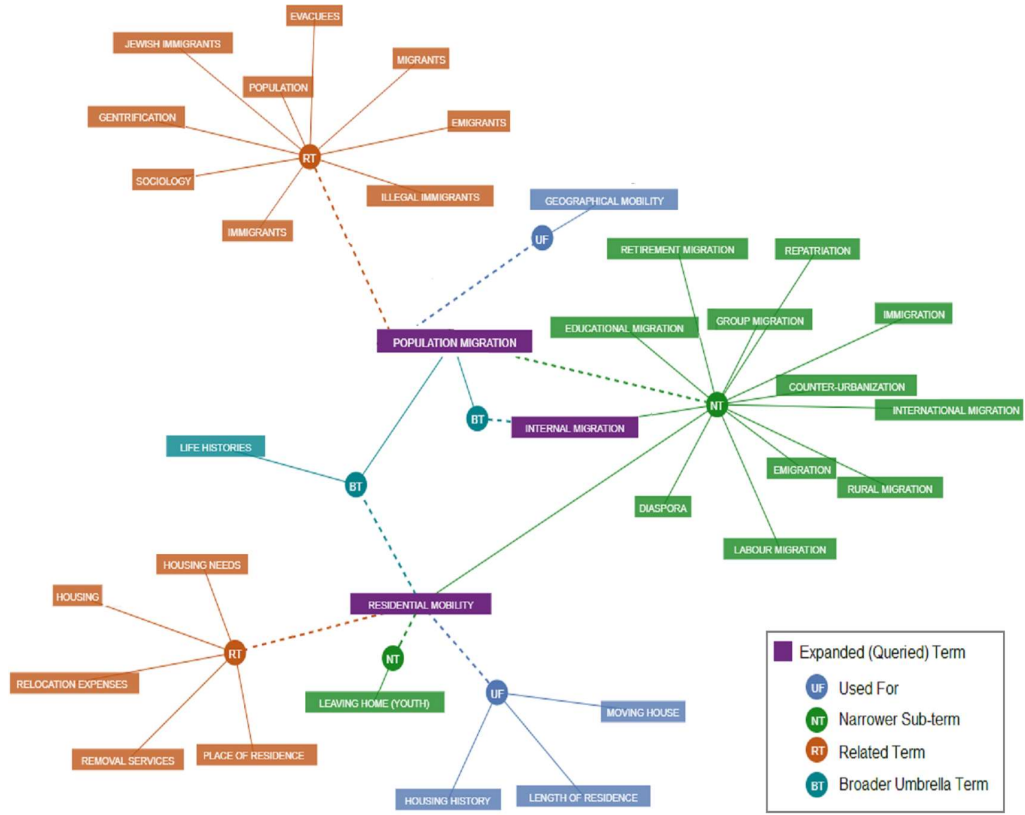

*Figure 1.5.        Graph of 'Residential Mobility' as a keyword in the Hasset Thesaurus.*

Note: Distances have been shortened to fit elements on the page.

Consulting the Humanities and Social Science Electronic Thesaurus (HASSET) of the UK Data Archive was useful for signposting keyword combinations and literature. It reassured the intuitions developed from unguided reading and created the queries in the next section.

### 1.5.2.2   Queries Overview

Migration literature in this analysis is captured in four theoretical nuances, or genre knowledge subdomains, as shown in Figure 1.6. The most open-ended query is 'internal migration housing'. It was expected that this would be dominated by anglophone case studies. However, the extent of this was unclear. Therefore, this query was intentionally kept broad.

The remaining three are: 'residential mobility' – a sociological nuance that implies individual-level studies; 'internal migration' – which is the preferred term in the British literature; and 'urban migration' – an equivalent term used especially in North American scholarship in the context of moves relating to urban housing. Urban migration was included because it was expected that the urban setting would be where internal migration and housing would interconnect. However, as urban migration translates to very different phenomena in the Global South vs North, this query was further narrowed down with thematic and geographic specifiers. 'Population migration' was not included because it has an ambiguous meaning that differs between the over-represented natural and less-represented social sciences within GS.

The three theory-related queries mentioned above are further narrowed down to zoom in and exhaust all relevant papers. This meant adding specifiers to the queries for the relevant context (Britain OR UK OR England), the dataset of interest ('census') and the housing subtheme ('housing market' was used where housing on its own is too broad to yield relevant results). The resulting query strings are shown in Figure 1.6.

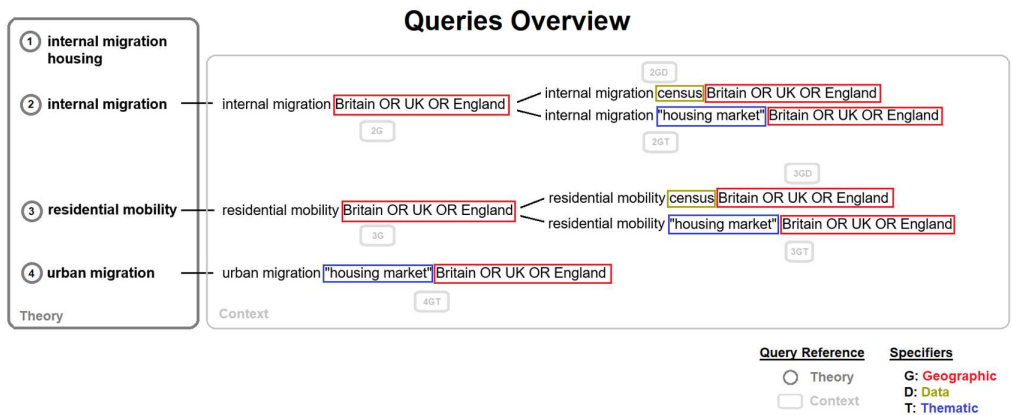

Figure 1.6.     Queries Overview

### 1.5.3  Manual Pruning

After the selection of queries, the second crucial step was the manual evaluation of papers in the scoping corpus – the corpus of documents gathered in the first scrape. This closer manual examination involved pruning 100 from 760 papers in the scoping corpus. The elimination of papers was done in three steps. The first step involved sifting through the titles only to split these into five groups, listed in order of interest: '4 - Look into' (n=107), '3 - Suitable' (n = 212), '2 - Check' (n=137), '1 - Marginal' (n = 81), '0 – Unlikely' (n = 223). The last three groups were pooled together for a more laborious evaluation which involved checking the abstract (n = 237); or full papers (n = 204). The focus was placed on the last three groups because it is more advantageous to remove irrelevant papers rather than miss potentially useful ones within the queries. For example, group '2 - Check' included case study papers that were included based on the quality of their literature review, specifically with reference to migration theory. In total, just under one-third of titles had their full text inspected during this manual evaluation.

After the set of eligible papers has been identified (N = 660), screened and pruned, as shown in Figure 1.2, these were used as the main 'seeds' for harvesting papers for the final corpus. The seed papers were complemented with four additional references of interest discussed in section 1.5.4., such that the total number of seed papers was N = 664. Where seed papers have been cited more than 100 times (as in the example in Figure 1.3-3), only the top hundred papers ranked by their 'Cited by' count are included. This yields a citation network that consists of the manually screened eligible papers and the most popular papers in these seedlings' academic lineage. Two layers were enough to gain wide coverage for two reasons: citation networks grew exponentially with each new paper, but also the references of interest included older papers, such as Ravenstein (1889).

### 1.5.4 References of Notable Interest

In addition to the queries in Figure 1.6, four key papers for the framing of this project are included as seeds for the final corpus. These are shown in Table 1-2. Although it is tempting to include more papers at this stage, references of interest were chosen sparingly. This was done because the exponential nature of citations means that each paper added could add up to 100 000 papers to the final corpus (100x100 papers for two citation layers, as shown in Figure 1.4).

The four papers included reflect a specific theme and are each a highly cited work chosen based on exploration during research for a narrative literature review. Ravenstein (1889) is included because this paper is widely considered the starting point of modern migration studies, but also the first of a lineage of quantitative population geography papers (as Ravenstein himself was a geographer!). Similarly important is Sjaastad (1962), who examines migration quantitatively at a macro level. However, unlike Ravenstein, Sjaastad takes a distinctly economic angle by framing migration in terms of resource allocation and income differentials. Sjaastad (1962) is also among the top five most important papers *both* within internal and international migration (Nestorowicz and Anacka, 2018, p. 295).

To complement the top-down theories, the final two references - Lee (1966) and Rossi (1955) - are included to reflect a micro-level approach important for understanding the characteristics of migrants. Lee (1966) extends the theoretical scholarship by introducing the idea of individual-specific 'intervening obstacles', which are essential to understand migration as a demographically differentiated experience. All three papers mentioned thus far are listed in this order as 'classic writings' on migration (King, 2020, p. 4) - a curious coincidence noticed in retrospect of data collection for this paper. The final reference of interest is Rossi (1955). This monograph is included as a starting point for sociological excursions into the role of the household unit over time for migration.

|     | *Title* | *Author* | *Year* | *Journal* | *Citations* |
| --- | --- | --- | --- | --- | --- |
| (1) | The Laws of Migration | EG Ravenstein | 1885 | Journal of the statistical society of London | 5012 |
| (2) | The costs and returns of human migration | LA Sjaastad | 1962 | Journal of Political Economy | 6057 |
| (3) | A Theory of Migration | ES Lee | 1966 | Demography | 6592 |
| (4) | Why families move | PH Rossi | 1980 | SAGE Publications | 1833 |

*Note: Cited by counts are correct as of August 2020*

*Table 1-2   Key Papers*

### 1.5.5 Natural Language Processing

Titles of documents were converted to lowercase, cleared of numbers and all punctuations except for hyphens within compound words. A Bag-of-Words approach was taken to identify the most frequent terms, where titles of documents were tokenised into individual terms and pooled together. Furthermore, to highlight the most important terms, 174 common English stop words from the SMART and Snowball lists were removed. These include words such as 'the', 'of',' from' and others (quanteda, 2019) which add little thematic insight.

A term document matrix representation was used to compute the correlation score between words or how likely two words are to co-occur within the same title. This is a binary matrix such that each document title is represented as a column, and all possible words within the corpus are listed as rows. Because this creates a very sparse matrix (213k unique terms across 445k documents), only terms occurring over fifty times were included, and words were stemmed using the Porter algorithm (Porter, 2006). Stemming involves normalising terms by removing common endings, such that only the root or 'stem' of the word remains, allowing for variants of a word to be grouped together. Lowering the resolutions of terms using this process was only useful for evaluating correlation scores. This is because different words of interest within this corpus have a common stem; specifically: 'house', 'housing' and 'household' would all be reduced to 'hous'.

An R implementation of Google's Compact Language Detector 2 was used to identify the language of titles. This is a probabilistic model that searches for tuples and triples combinations that are characteristic of a language to identify one of 83 languages (CLD2Owners, 2021). Although his algorithm is superseded by a neural network model (CLD3), the earlier implementation was found to be more successful by providing a coverage of over 90% successful translations of titles within the citation corpus of 445k documents.

## 1.6 Results:

### 1.6.1 Scoping Titles

#### 1.6.1.1 Top Titles

According to Google Scholar's proprietary but openly available ranking algorithm, the scraping technique returns a scoop of the first 100 relevant papers. To use a metaphor: the returned top one hundred articles are only the tip of a mountain of documents related to the search terms. The top title per query is shown below.

| Title | Year | 1_IMH | 2_IM 2G | 2_IM 2GD | 2_IM 2GT | 3_RM 3G | 3_RM 3GD | 3_RM 3GT | 4_UM 4GT | Authors | Cited |
|---|---|---|---|---|---|---|---|---|---|---|---|
| Does land use planning shape regional economies? A simultaneous analysis of housing supply, internal migration and local employment growth in the Netherlands | 2009 | 1 | | | • | | | | | Vermeulen and Ommeren | 63 |
| Internal migration and ethnic groups: evidence for Britain from the 2001 Census | 2008 | • | • | 3 | 1 | 2 | | | | Finney and Simpson | 132 |
| Population movement within the UK | 2005 | | | • | • | 1 | | | | Champion | 117 |
| Intra-urban migration and housing submarkets: Theory and evidence | 2004 | | | | | | | • | 1 | Jones and Leishman | 99 |
| Migration in a mature economy: emigration and internal migration in England and Wales 1861-1900 | 2003 | | • | 1 | • | | | | | Baines | 396 |
| Tied down or room to move? Investigating the relationships between housing tenure, employment status and residential mobility in Britain | 2002 | | | | • | • | 1 | 2 | • | Böheim and Taylor | 241 |
| Internal migration in developed countries | 1997 | 1 | • | | | | | | | Greenwood | 1090 |
| Spatial mobility, tenure mobility, and emerging social divisions in the UK housing market | 1987 | | | | | | | 1 | | Forrest | 82 |
| Residential satisfaction as an intervening variable in residential mobility | 1974 | | | | | 1 | | | | Speare | 876 |
| Life-cycle, housing tenure and intra-urban residential mobility: A causal model | 1973 | | | | | • | 1 | • | • | Pickvance | 86 |
| The intra-urban migration process: a perspective | 1970 | | | | | | | | 1 | Brown and Moore | 1223 |

Note: Cited by counts are correct as of August 2020; Ranks 1-3 included otherwise shown as •

*Table 1-3.        Top Titles per Individual Query*

Figure 1.6 shows that queries achieve a narrow thematic scope as the top documents correspond to more than one query. It also shows that the oldest paper (1970) and top result for 'urban migration' (4_UM) has the highest citation count with 1 223 citations. Furthermore, the theoretical queries from Figure 1.6 are almost all at least one order of magnitude above the top results when these are narrowed down by geographic context. The only exception is 'internal migration housing' (1_IMH). The top entry for this is both the newest paper in the list, and the least cited one, suggesting that the scholarship on internal migration coupled with housing is a comparatively younger subfield.

### 1.6.1.2 Pruning Papers

Whereas the previous section described the typology of relevant papers, this section will discuss which papers were less relevant and therefore discarded before commencing the citation collection. Each of the eleven queries lost papers during the manual pruning of the scoping corpus (n=760). The papers removed fell broadly into three groups. These included: medical papers (e.g., 'Selection bias from differential residential mobility as an explanation for associations of wire codes with childhood cancer'), descriptive foreign case study too specific for comparisons with the UK (e.g. 'Internal migration in the Klang Valley of Malaysia: Issues and implications') and historic studies pre 1800 (e.g. 'What's in a settlement? Domestic practice and residential mobility in Early Bronze Age southern England '). The stepwise methodology of pruning papers at this stage is discussed in section 1.5.3.

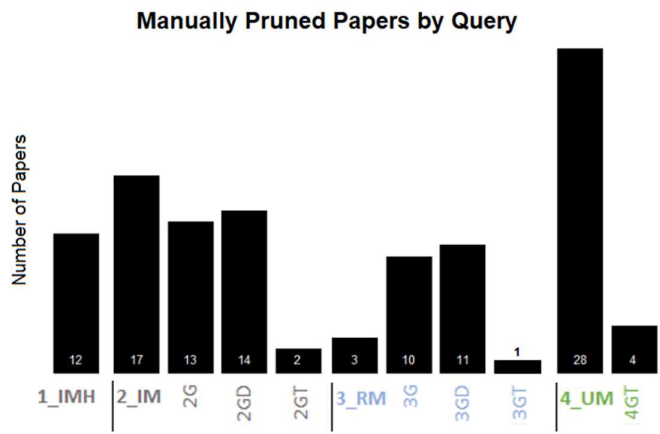

Figure 1.7    Manually Pruned Papers by Query Group

As expected, Figure 1.7 shows that the broadest search queries -i.e. those relating to theories - had the largest amount of tangential papers. Of these, UM had the largest volume of omissions. This keyword pairing is often associated with the literature of case studies in developing countries. By contrast, close to all the papers returned by combining both 'housing' as a theme (T) and the relevant geographical specifiers (G) were retained.

Interestingly 'residential mobility' as a search term is associated with a handful of historic papers using isotope analysis to measure migration (e.g. 'Anglo-Saxon residential mobility at West Heslerton, North Yorkshire, UK from combined O-and Sr-isotope analysis' published in an edited volume on plasma mass spectrometry). The most entertaining outlier title removed in this stage was one from the field of ethnomusicology entitled "Moving away from silence: Music of the Peruvian Altiplano and the Experience of Urban Migration". A total of one hundred papers were methodically removed at this stage (as discussed in greater detail in section 1.5.3), leaving N = 660 papers in the seed corpus of papers.

### 1.6.1.3 Four Thematic Genres

Different disciplines may use language differently to give concepts meaning and frame research. Jargons of genre knowledge can then delineate epistemic communities and create discipline-specific enclaves for migration scholarship to be advanced. This means that different thematic genres of migration scholarship need to be considered for an exhaustive identification of relevant papers. The thematic genres used to create the scoping corpus correspond to the four theory groups in Figure 1.6.

Figure 1.8 is a Venn diagram of seed papers returned by the four broadest theoretical queries, as shown in Figure 1.6. It clearly shows that the most popular document results on residential mobility form a distinct body of scholarship relative to internal migration and urban migration. It also shows that the housing thematic features only in a small subset of the most popular and influential (by citation count) papers on IM: of the top hundred documents under IM, only 25 papers (25%) include some reference to housing. Finally, Figure 1.8 shows that UM, similarly to RM, is less quickly associated with the epistemic communities around internal migration.

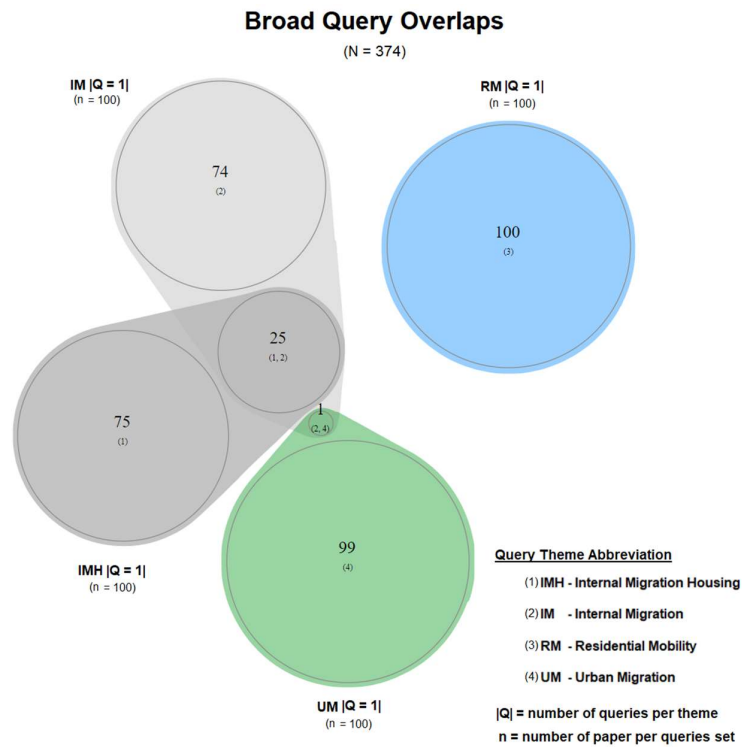

*Figure 1.8.     Seed Results of the Key Search Terms*

Most importantly, Figure 1.8 shows that the RM and UM lens on mobility is distant from the general body of scholarship on internal migration. Distant, however, does not mean distinct. This is because if one considers enough papers, the exponential nature of citations means that any two papers will eventually become related.

### 1.6.1.4  Overlap of Thematic Genres

To collect papers at the intersection of the genres discussed in section 1.6.1.1, these were further narrowed down by geography, data and theme, as shown in Figure 1.6. Noteworthy is that the unique set of papers in the scope corpus is 760 papers - as opposed to 1100 titles - which would have been the case if each of the eleven queries returned one hundred mutually exclusive papers.

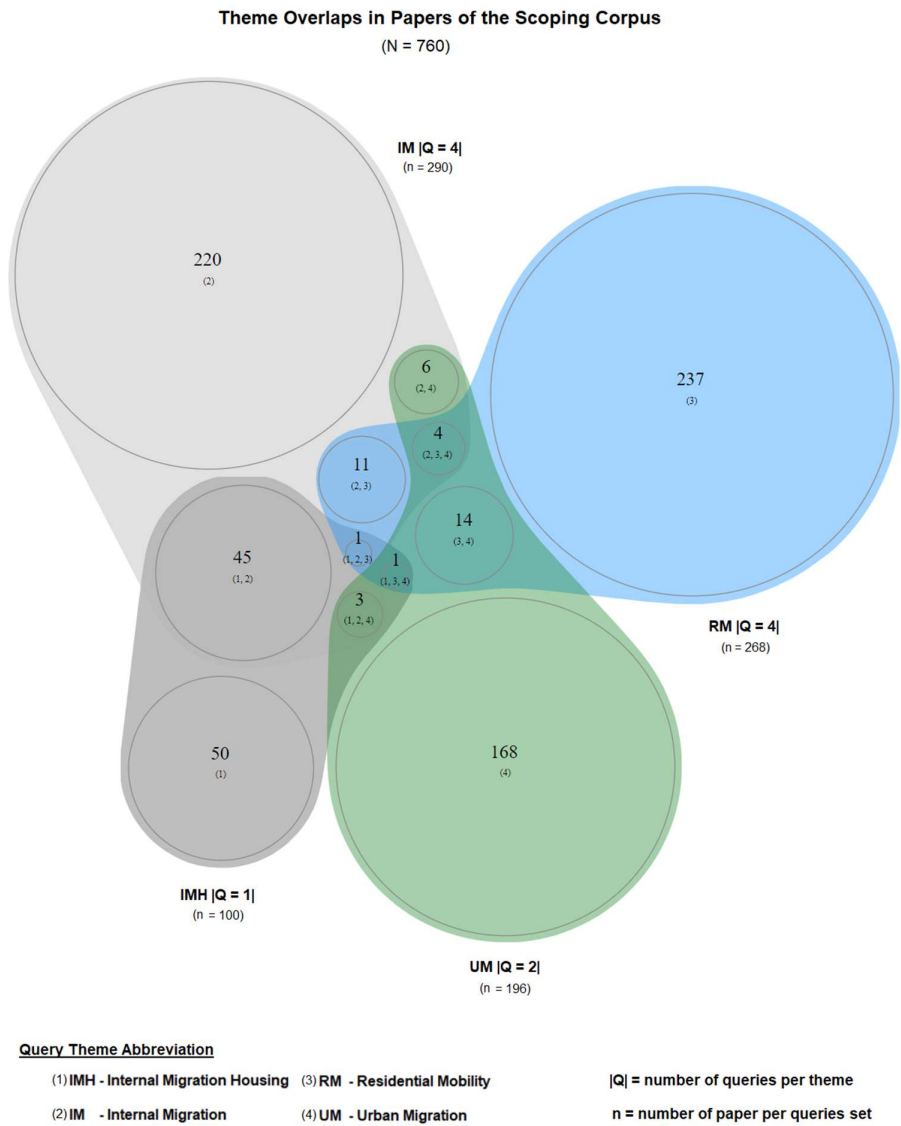

Figure 1.9     Seed Results Overlap of Paper Themes

The purpose of having eleven queries was to be exhaustive. However, when examining eleven groups, this is unwieldy; therefore, these are shown by their thematic genres. The total scoping corpus by theory genre is shown in Figure 1.9. This mirrors the results shown in Figure 1.8: even with the additional query specifiers, the four epistemic communities are distant as measured by their most popular papers. The Venn diagram depicts the number of papers that are covered by more than one query group. Each paper can co-occur in more than one query – i.e. it can come up as one of the top results in more than one GS search. Should IM and IHM be considered jointly, there are only 40/760 papers – 5% of documents - that overlap between theory themes.

### 1.6.1.5 Query Overlaps

Figure 1.10 takes a closer look at how papers in the seed corpus overlap across all eleven queries by thematic genre. Given all papers, each paper can be thought of as an element in intersecting query sets. However, as eleven sets are hard to see and make sense of in a Venn diagram, the intersection of queries is shown as the columns of Figure 1.10. The rows in Figure 1.10 are query strings.

At the top of each column is a bar chart showing the total number of papers that match that specific intersection pattern of queries (this is equivalent to the surface area in a Venn diagram). Query rows are grouped by their theoretical framing (as per Figure 1.6) into boxes. Therefore, each row is either one of four theoretical query abbreviations (shown as stars) or a version of these with narrower search specifiers (Figure 1.10, bottom right).

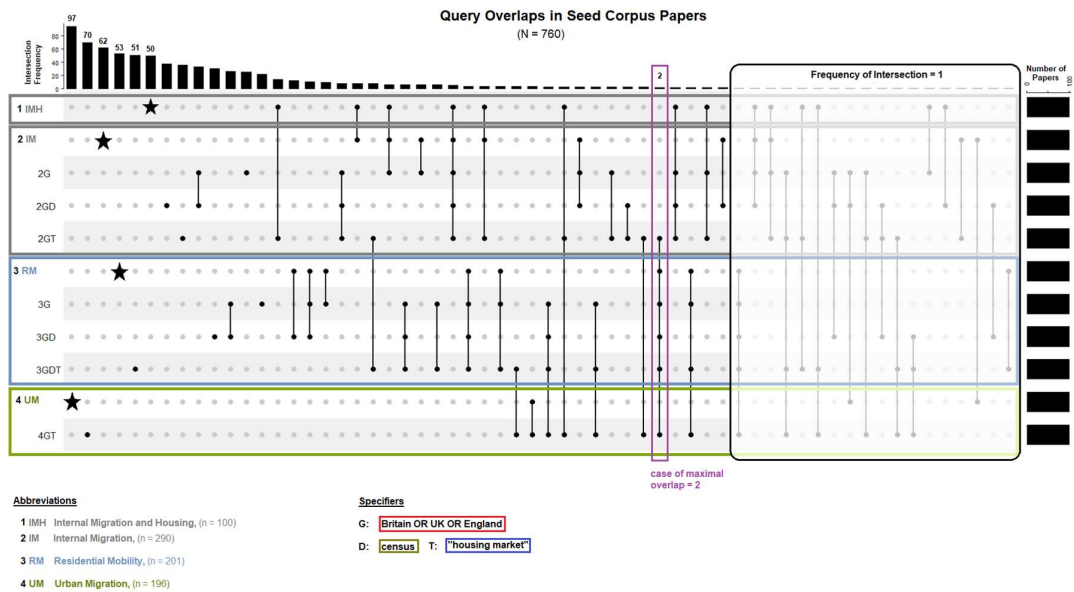

Figure 1.10.    Query Intersections by Paper Frequency

The first thing to notice about Figure 1.10 is that, as expected, the four broadest theoretical queries (IHM, RM, IM and UM – shown in stars) return the largest number of unique papers. Notable, however, is that 'urban migration' (UM) is the most distinct of all (97/100 unique papers) – capturing the anticipated split in meaning associated with the phrase for the literature of the Global North (to mean within city moves) and South (to mean moves to urban areas). The next largest distinct group of papers were those unique to 'internal migration' (IM – 70/100 unique papers). This is the case because IM is - by definition - bound to national boundaries, which may not include or engage with a comparison to the UK context. Noteworthy, too is that the maximal overlap is six queries for two papers (shown in purple in Figure 1.10) – both cover all queries related to RM.

### 1.6.1.6 Papers of Maximal Overlap

Figure 1.11 shows a closer look at the amount of overlap between the queries showing that a double overlap is most common. This is due to the exhaustive approach taken. The double overlap corresponds to a match on a 'theoretical query' as shown in Figure 1.6, and the same theoretical query narrowed down to include engagement or comparison with the UK or census data. Table 1-4 lists the papers that were overlapped.

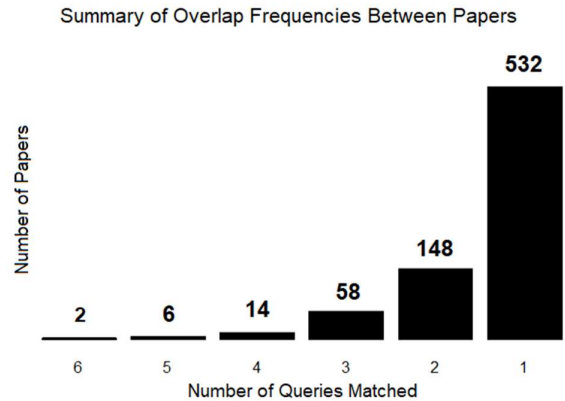

Figure 1.11.    Overlap of Eleven Queries

| Titles | Hits | Year | Cited by | Author |
|---|---|---|---|---|
| The definition of housing market areas and strategic planning | 6 | 2002 | 113 | Jones |
| Moving house, creating home: Exploring residential mobility | 6 | 2002 | 182 | Winstanley et al. |
| Intercensal mobility in a Victorian city | 5 | 1977 | 44 | Dennis |
| In and out of Chinatown: Residential mobility and segregation of New York City's Chinese | 5 | 1991 | 173 | Zhou and Logan |
| Does land use planning shape regional economies? A simultaneous analysis of housing supply, internal migration and local employment growth in the Netherlands | 5 | 2009 | 63 | Vermeulen and Ommeren |
| Interregional migration and housing structure in an East European transition country: A view of Lithuania 2001-2008 | 5 | 2009 | 29 | Bloze |
| Ethnic segregation and residential mobility: relocations of minority ethnic groups in the Netherlands | 5 | 2010 | 138 | Bolt and Kempen |
| Explanations for long-distance counter-urban migration into fringe areas in Denmark | 5 | 2011 | 43 | Andersen |

Table 1-4    Papers with Maximal Overlap

Table 1-4 shows that few papers are covered by many queries simultaneously. For example, Jones (2002) – whilst presenting an empirical analysis of Glasgow - argues that the key variable in the geographic delineation of housing market areas should be 'migration self-containment' (2002, p. 549). Importantly he provides a review of how housing market areas are understood and defined differently by academics - who might advise on policy - and planners - who translate policy guidance into decisions on the ground. By contrast, Winstanley et al. (2002) take a qualitative ethnographic approach based on interviews of families in Christchurch, New Zealand, to unpack themes of 'house' and 'home' over time, highlighting the complexity

of mobility decisions. This paper uses anecdotes following the scholarship of Rossi (1955) to discuss adjustment moves as a function of family needs on housing, thus acknowledging a plurality of possible life course trajectories.

The papers in Table 1-4 show the full spectrum of how a place can be taken into account. There are three broad approaches. Firstly, Jones (2002) – like Vermeulen and Ommeren (2009), Bloze (2009) and Andersen (2011) - take an ecological, macro approach within a spatial nuance through empirical studies of specific, named geographies. By contrast, Winstanley et al. (2002) – like Zhou and Logan (1991) – present the anecdotal, interview-based approach where the prominence of places is subjective to the individual experience. These are the two polar approaches for representing the role of geography: by focusing either on 'named' places and thus giving greater insight on the characteristics of a place, or 'named' people and thus giving greater insight into the motives of individuals.

Finally, the midway path of linking place to social processes is to use it as a category. This requires a differencing metric that can be applied to the built environment to delineate geographies based on their features into places that are broadly 'the same' or 'different' based on a generalised description of a place. In this approach, the focus is no longer on 'named' places or people but rather on 'types' of places and people. This midway approach is often taken using feature-rich census or household panel data, as is the case with Dennis (1977) and Bolt and van Kempen (2010), respectively.

### 1.6.2 Seed Titles

#### 1.6.2.1 Themes and their subqueries

Assuming a grossly conservative Fermi estimate of reading one document within 22 minutes, as derived in section 0, the seed corpus of N = 660 papers would take close to 7 weeks to 'read' at a rate of 37 hours/week. Therefore, to give a more efficient sense of the seed corpus, this section provides a short overview of the key themes based on the wording of document titles.

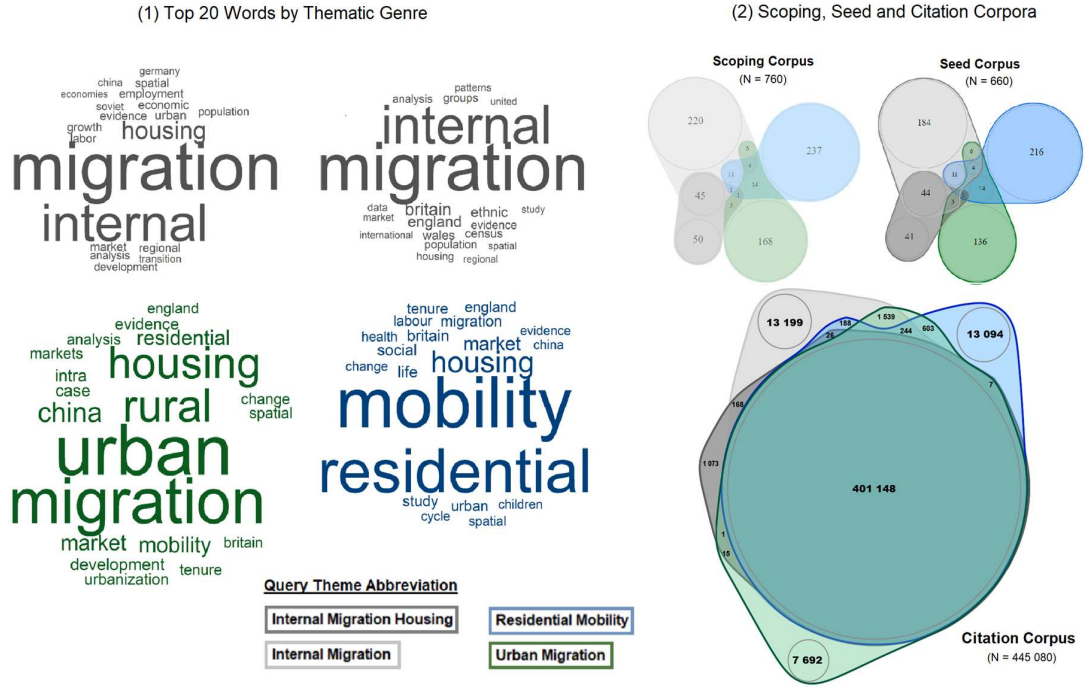

*Figure 1.12     Overview of the Four Thematic Genres*

Figure 1.12 provides an overview of the four thematic genres as keywords (1) and as represented by papers at each stage of creating a corpus of relevant papers (2). The Venn diagrams illustrate the corpora at each step, most notably the final citation corpus of papers for which there is a greater overlap of the thematic genres. The top twenty keywords from the titles of papers within the seed corpus are shown where the size of each term depends on its frequency. The first thing to note is that all themes include housing (because this was specified by the queries shown in Figure 1.6.); however the prominence of the term differs most notably between IHM and IM. The IHM scholarship shows a nuance towards studies of the former Eastern European bloc with words such as 'soviet', 'transition' and 'Germany'. This focus is likely because the transitions of these places from a socialist to a democratic model of governance included the radical shift from almost exclusively state-owned housing to a privatisation thereof in conjunction with free movement after the fall of the Iron Curtain. IM, on the other hand - without the forced focus on housing - features top keywords such as 'Britain', 'England' with 'ethnic', 'groups' and 'population' as themes. This suggests that the focus of the IM linage of papers lies in the Anglo-centric case studies of segregation. As expected, RM introduces a focus on individual-level household and family themes through terms such as 'life', 'cycle', 'children' and 'social'. By contrast, UM ushers in macro spatial themes with terms such as 'rural', 'development', 'change' and 'urbanisation'.

### 1.6.3 Citation Corpus

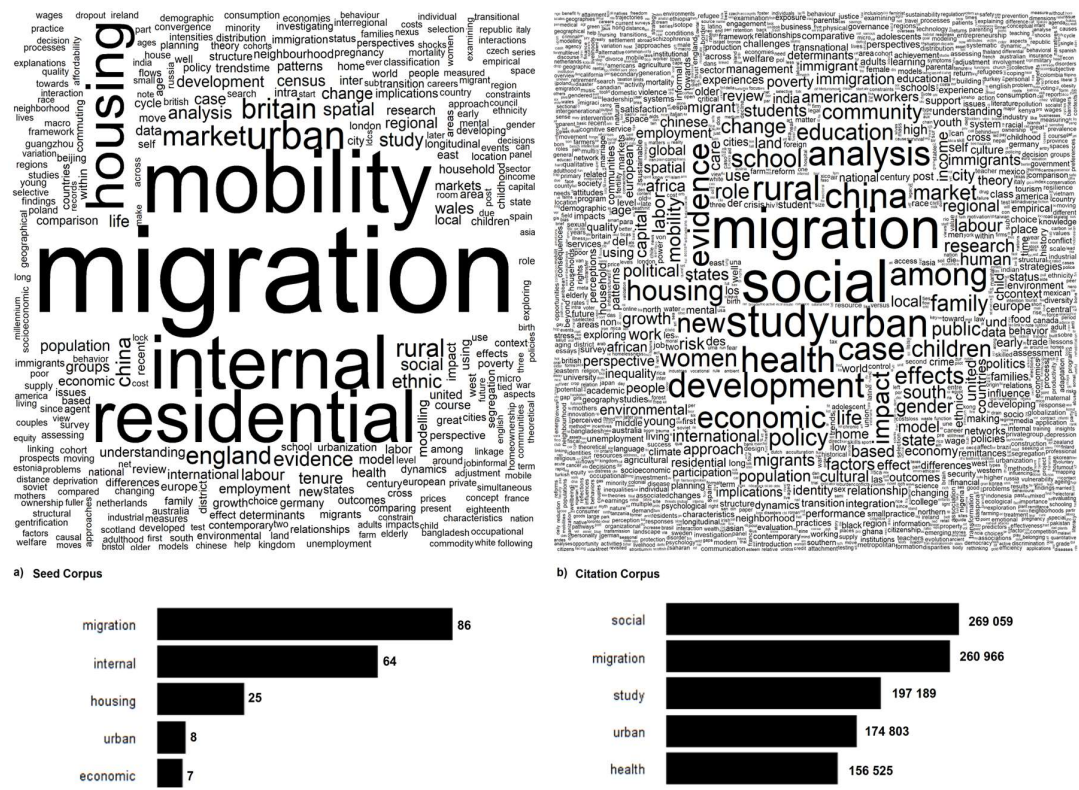

*Figure 1.13    Top 15 Words used in paper titles*

Figure 1.13 juxtaposes the seed and citation corpora and examines the top five most prominent terms in the titles of each set of papers. The citation corpus is much larger as it consists of the seed corpus and the top hundred most popular papers citing these up to two steps (as illustrated in Figure 1.4.). The is why it has many more terms, as seen in Figure 1.13-b). Noteworthy in this juxtaposition is the prominence of 'social' over 'migration' as a term within the citation corpus. This suggests that migration and housing research is cited widely beyond the epistemic communities of its researchers to help understand other social processes. The essence of these processes can be seen by examining which terms' social' is most associated with within document titles. Figure 1.14-a) suggests that there is a precedent of relating migration to specific social processes, notably to do with individual-level benefits, such as social capital, network(s) and support; and to a lesser extent, processes that might manifest themselves across space such as exclusion and cohesion.

Noteworthy from Figure 1.13-b) is the lack of 'housing' as a top term. This is because it only ranks 15[th] in popularity within the citation corpus. 'Housing' is less prevalent than 'health' (ranked 5[th]), but with a frequency within the same order of magnitude as 'economic' (ranked 12[th] with 132k mentions). This is unsurprising when the term is examined closer, as shown in Figure 1.14-b): housing has a theme features prominently for its market nuance to do with affordability, price, and to a smaller extent tenure. The relatively low correlation scores for both terms - indicative of how likely pairs of words co-occur - reflect the diversity of terms as well as the breadth and plurality of themes these are used in conjunction with.

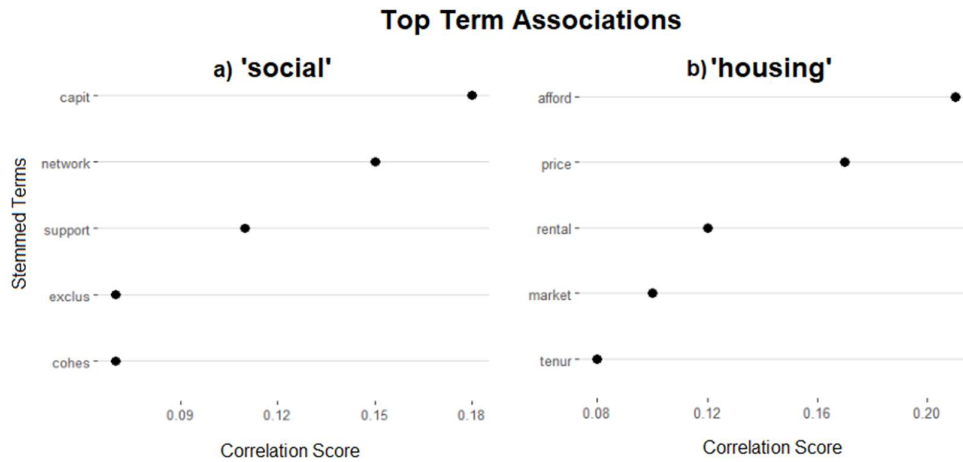

*Figure 1.14    Stemmed term associations for 'social' and 'housing'*

### 1.6.3.1  Internationalisation

Figure 1.15 shows that the corpus of papers collected for this systematised review increased exponentially by three orders of magnitude across all years. Noteworthy is also the fact that papers in languages other than English feature with almost consistently increasing prominence and account for 17% of documents across the corpus. This is a clear indication that scholarship at the intersection of migration and housing have a global scope.

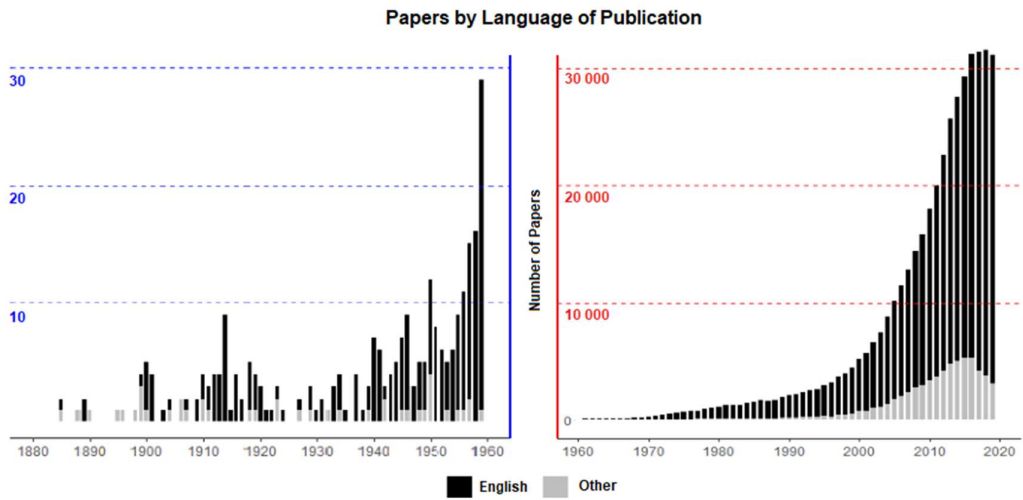

*Figure 1.15.    Papers by Language of Publication Across Time*

Interestingly, the second most prominent language of relevant research is simplified Chinese (3% of papers). 'China' as a term is prominent in both the most top hundred papers (shown in Figure 1.12) and the final corpus, where it ranks 6[th] in word frequency. This is to be expected, given that China is the single largest country in terms of population count – making up for just over 18% of the world population in 2020.

### 1.6.3.2 Bridging Documents

Although the networks are overall sparse, not all documents within these are of equal importance. Documents with a high betweenness centrality are those that commonly appear on the shortest path between other documents within the corpora. These are important documents because they serve as bridges that keep the body of scholarship connected. The top bridging papers for the final corpus are shown in Figure 1.16, and the equivalent for the housing subnetwork is shown in Figure 1.17.

It is worth noting that the primary bridging documents by a large margin in both networks are books. Although there are three clear bridges within the citation network of housing, the full corpus of papers examined has several highly connected nodes. Figure 1.16 shows the multi-edition book on international migration by de Hass et al. (2020) to be the most important document by a considerable distance. It argues that as of the 1990s, migration – notably international migration – may be the single most important factor of global change with socio-political implications. This argument is revised and refined across the different editions between 1993 and 2020 to reflect emerging concerns such as the implications of the 2008 economic crisis as well as a broader discussion of non-Western regions. *The Age of Migration* is followed in rank by other books and edited volumes that engage with international moves, the rural-urban distinction and development more broadly.

Two papers stand out against the older, longer documents. These are Champion (1994) in *Urban Studies* and Schewel (2019) in *International Migration Review*. Champion (1994) argues for a *demographic* lens to interpreting international migration. He thus urges efforts towards harmonised statistics on the topic, which resonate almost three decades later as migration data is still a contentious issue. The only other journal article that emerges as an important bridging document in Figure 1.16 is Schewel's (2019) positional paper on migration's flip side – immobility. She argues that the capability and aspiration to move or stay may lie on a continuum, such that diverse methodological strategies could give indications of stability by capturing individuals' centres of gravity over time. Importantly, this paper argues that staying and moving are are equally important responses to wider structural forces. This conceptual framing of mobility and immobility is embedded in Chapter 6 of this thesis.

Examining the housing subset (Figure 1.17) shows a triad of bridging documents. Unlike the larger corpus, only the top reference is a book with two papers ranked second and third. The top bridging papers are Dietz and Haurin's (2003) agenda-setting review of homeownership and Clark et al.'s (2003) US case study, which suggests a mostly upward trend in quality, price and tenure for adults between 1968-93. The lack of monographs in the top bridging documents might in itself suggest the under-developed nature of housing within migration research. Indeed, only the very top document is a book first published in 1998 and again in 2012 by Prof. William A. V. Clark in partnership with Prof. Frans Dieleman (Clark and Dieleman, 2012), which focuses on household choice and outcomes in the US and Dutch contexts. The importance of their monograph is its ability to link households to features of the housing stock such as tenure and size (or housing hierarchies as referred to in their book). It suggests that households provide an important lens to understanding migration and housing outcomes in shifting socio-economic and political realities.

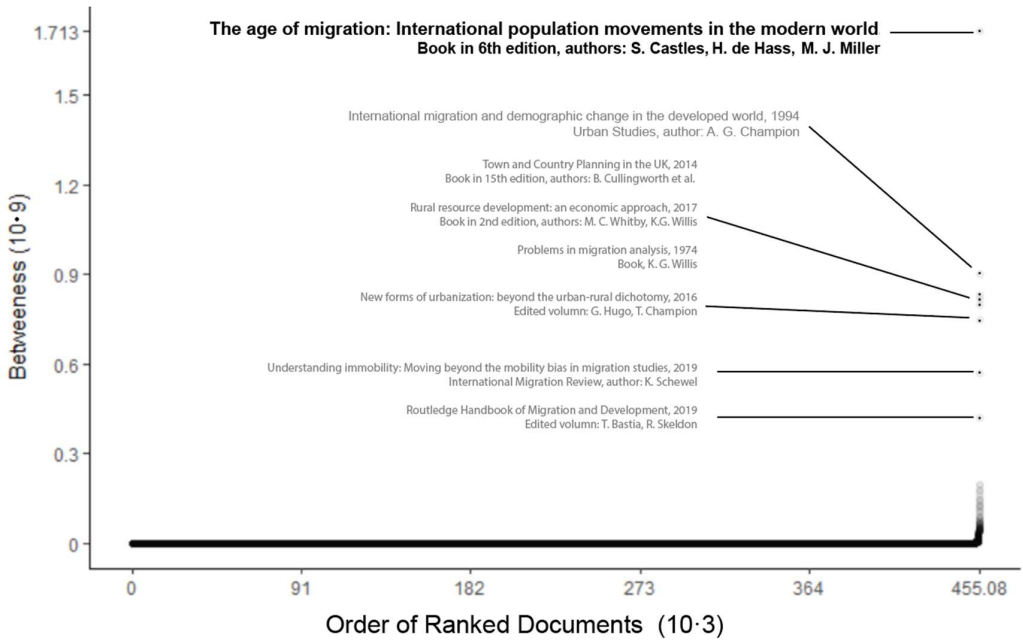

Figure 1.16   Bridging documents of the Final Corpus

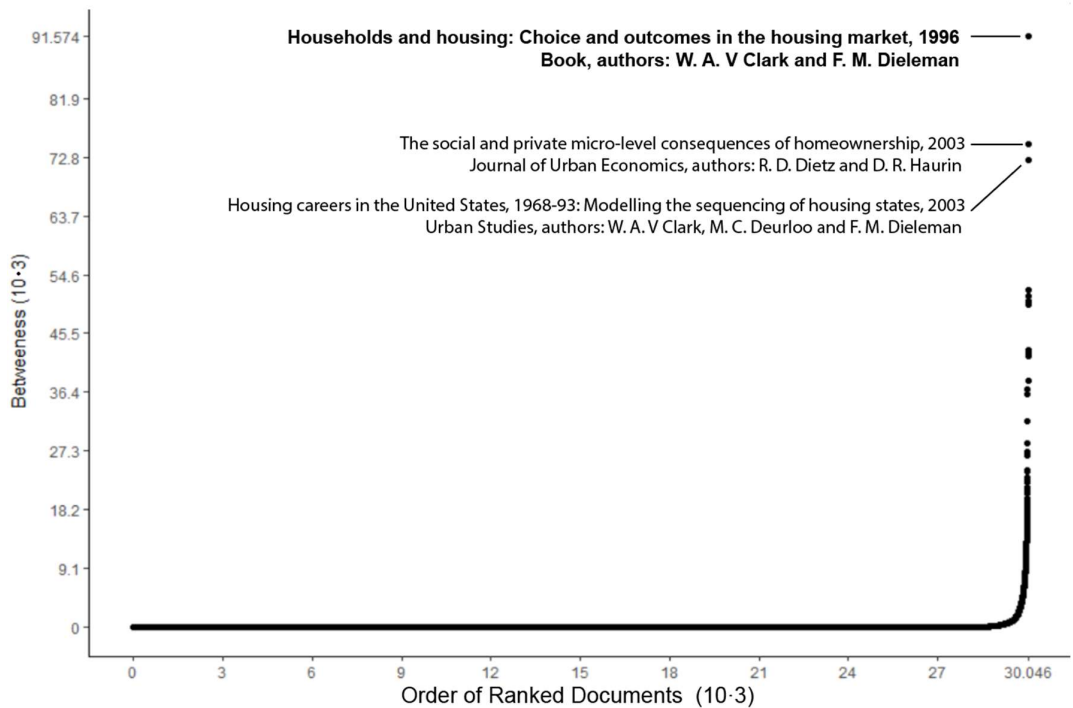

Figure 1.17   Bridging documents of the Housing Subset

## 1.7 Limitations

This analysis situated documents on the scholarship intersecting migration and housing. It reflected on and quantified the links between the two fields. Despite providing a global overview, citation rates are a metric to be taken with a grain of salt for several reasons. Occasionally, papers of contest quality may be highly cited because the quality of their output journal in highly regarded compared to a similar paper in a new journal (Richards *et al.*, 2009, pp. 7–10). More importantly, however, citations are retrospective. They will imminently give older documents an advantage over newer ones. Furthermore, the popularity of documents is also dependent on differing indexing standards. Notably, within-discipline indexing provides an advantage with knowledge accumulation within pre-defined key terms and concepts that are easier to find. This is advantageous in the medical sciences, notably for controlled clinical trials, but more difficult to keep pace with with regards to evolving social paradigms.

## 1.8 Discussion

This paper presented a systematised review of the broad body of scholarship on migration and housing gathered using Google Scholar (GS) – the most extensive bibliographic database due to its inclusion of a range of sources (Gusenbauer, 2019; Martín-Martín, Thelwall and López-Cózar, 2020). The analysis used four thematic genres (Figure 1.12), each with a different angle to framing the synergetic processes between the built environment and its inhabitants through migration. Examining the documents at intersections of several queries in section 1.6.1.6 showed a triad of typologies discussing the role of place. These typologies include:

- a descriptive approach on the characteristics of 'named' places
- a causal/subjective approach based on experiences or judgments of places by 'named' persons
- and a contrasting approach focusing on 'types' of places

Examining the final corpus of papers suggested that most documents were empirically driven case studies: the third most popular key term in Figure 1.13-b) is 'study'. The prominence of studies is noteworthy because it reaffirms that literature at the intersection of migration and housing relies on empirical case studies and inductive reasoning. Such studies are observations of reality that, at their best, generalise to create theories and paradigms that interlink to stand the tests of time when a topic area matures. As the scholarship around migration and housing spans many disciplines, the findings of this paper suggest understanding migration and housing is a task of grafting diverse ideas into a shared, integrative framework.

Citation networks are useful for a broad identification of the 'essence of the field' (Foster *et al.*, 2007, p. 3), where the amount of scholarly output pushes the limits of human reading time. Given the exponential increase in migration-related research output, as confirmed in this analysis and observed by Pisarevskaya et al. (2020, p. 463), a case study's ability to generalise is increasingly urgent, especially so with the growth of international case studies and the second wave of comparative scholarship these triggers. Across the corpus, 17% of the top articles returned by the search method are not in English. Furthermore, not all publications that are in English refer to the English context. Therefore, the percentage cited is a lower estimate of research beyond the dominant British and American contexts. Also, papers included within the corpus analyse are only the highly cited ones in GS, and consequently, unlikely to represent the actual volumes of research activity at a world scale. Replicable research – one that can be generalised across places - is therefore essential. It may even taking precedent as a contribution over traditional descriptive or anecdotal accounts in the age of data and computing, where we drowning in data but thirsty for understanding.

Finally, examining the final corpus of papers linked through citations shows the broad spectrum of documents and the plurality of themes related to migration and housing. This dispersion is consistent with previous observations on the breadth of the thematic: examining publication data in the 'Geography and Environmental Studies' category from the Research Assessment Exercise (RAE) of 2008 (N= 4 590 written outputs), Richards et al. reach the same conclusion. They note: 'both physical and human geographers are promiscuous in their choice of journals' (2009, p. 234).

## 1.9 Conclusions

Though bibliometric analysis may identify and assess retrospective 'impact' through citation counts, the manual peer review may be a closer approximation of contemporary perceptions of 'quality'. However, both reviews and bibliometric analysis are examples of engagement which drive the epistemic endeavour. Journal articles – the most widely indexed content in specialised academic databases - are not the only, or in fact, the universally best form of knowledge dissemination. Books are essential for the collaboration and curation these undergo to 'produce a benchmark analysis of considerable influence and longevity' (Richards *et al.*, 2009, p. 3). However, more timely output such as working papers or blogs, white or policy papers, and teaching may have a much broader reach and more profound contribution to the development of a field.

This analysis examined key term overlaps (Figure 1.12, Figure 1.13, and Figure 1.14) to show research themes in the literature intersecting migration and housing. Importantly, it situated the scholarship on migration and housing, thus reflecting on the quantification of links between the two. Examining how documents are interrelated, this paper highlighted the relative under-engagement of the housing theme within migration research. Case studies formed most documents in the bibliographic dataset of 445 080 records examined in this paper. Such studies are observations of reality that, at their best, generalise to create theories and paradigms that strengthen ties between research output as topic areas mature. Considering the exponential growth in documents over time, a guiding principle for future contributions to the intersection of migration and housing would include *integrative* rather than *additive* approaches that scale across different contexts. In this way, the multitude of emerging datasets in these domains may better serve as bridges across a sparse and sprawling interdisciplinary landscape.